\documentclass[aps,prd,preprint,groupedaddress]{revtex4}
\usepackage{bm}
\usepackage{graphicx}
\usepackage{amsmath}
\usepackage{amsfonts}
\usepackage{amssymb}

\begin{document}

\title{Quantum Interference to Measure Spacetime Curvature: A Proposed
  Experiment at the Intersection of Quantum Mechanics and General
  Relativity}

\author{Raymond Y.~Chiao}
\email{chiao@socrates.berkeley.edu}
\homepage[]{physics.berkeley.edu/research/chiao/}
\thanks{}
\affiliation{Department of Physics, University of California,
  Berkeley, CA 94720-7300}

\author{Achilles D.~Speliotopoulos}
\email{adspelio@uclink.berkeley.edu}
\homepage{}
\thanks{}
\affiliation{Department of Physics, University of California,
  Berkeley, CA 94720-7300}

\date{April 7, 2003}

\begin{abstract}
An experiment in Low Earth Orbit (LEO) is proposed to measure components of
the Riemann curvature tensor using atom interferometry. We show that the
difference in the quantum phase $\Delta\phi$ of an atom that can
travel along two intersecting geodesics is given by $mR_{0i0j}/\hbar$
times the spacetime volume contained within the geodesics. Our
expression for $\Delta\phi$ also holds for gravitational waves in the
long wavelength limit. 

\vskip24pt

\noindent{Keywords: Riemann Curvature Tensor, Atom Interferometry}
\end{abstract}

\maketitle

In general relativity no ``gravitational force'' exists \cite{MTW}; 
there is only geometry. Thus, in the
absence of any forces, a test particle will undergo free fall,
and will travel along a geodesic determined by the local spacetime
geometry. What we normally associate with the ``force of gravity'' on a 
particle in Newtonian mechanics is not a force at all: The particle is
simply traveling along the ``straightest'' possible path in a curved
spacetime.

Nowhere is this basic difference between Newtonian gravity and general
relativity more apparent than for tidal forces. Consider twin
astronauts in separate, but nearby, orbits around the Earth. In Newtonian
gravity the astronauts accelerate towards or away from one
another due to the \textit{difference} in the Earth's gravity between
them. In general relativity, by contrast, because both astronauts do
not feel any forces, they undergo free fall; they
travel along geodesics in a spacetime curved by the Earth. Because the two 
geodesics are slightly different, one of the astronauts
will see his twin drift toward or away from him. The \textit{rate} of
change of the distance $X^{\mu}$ separating the two astronauts depends
on the local curvature tensor $R_{\mu\nu\alpha\beta}$ \textit{along}
$X^\mu$. If $R_{\mu\nu\alpha\beta}$ changes slowly along $X^{\mu}$,
$X^{\mu}$ satisfies the geodesic deviation equation \cite{MTW}; if
$R_{\mu\nu\alpha\beta}$ changes appreciably along $X^{\mu}$, its
evolution depends on the \textit{integral} of $R_{\mu\nu\alpha\beta}$
along $X^{\mu}$ \cite{GLF}. In either case, one astronaut interprets
the shift in his twin's position as being due to a fictitious
Newtonian ``gravitational tidal force'' acting on his twin, even
though the latter feels no forces.

The twins age differently due to their differing gravitational
red shifts: The twin in the lower orbit ages
less than the farther twin. However, in this general relativistic twin
paradox, \textit{both} twins feel no forces, unlike in the special
relativistic case. Their age difference cannot arise within a Newtonian
framework, where time is absolute. However, at the quantum level, this
\textit{age} difference for interfering atoms becomes a measureable 
\textit{phase} difference directly related to the curvature.

In this Gravity Research Foundation essay, we propose an experiment in LEO
to measure components of the Riemann curvature tensor. In a quantum
extension of the above example, this experiment will test quantum
mechanically the tenant of general relativity that local geometry
determines motion. However, instead of relying on \textit{classical}
test particles, we make use of a quantum phenomenon: the fringe shifts 
of an interference pattern caused by phase differences of \textit{quantum}
test particles. Consequently, this experiment will probe the intersection of 
general relativity and quantum mechanics \cite{Chiao}. 

The conceptual roots of this proposed experiment are well established
\cite{Stod, Anan, GLF}. Its technical roots lie in Chu's \cite{PCC}
and Kasevich's \cite{MFFSK} work on atom interferometry 
in their measurements of the local acceleration due to gravity
$g$ using Mach-Zehnder-type interferometers
\cite{CarMly, KETP, RKWHB, KasChu, ROBSZ}. An accuracy based on
Newtonian dynamics of $\Delta g/g\sim10^{-8}$
for a single measurement, and up to $10^{-11}$ after a two-day
integration, was achieved. In these measurements, cesium atoms were thrown 
upwards in the Earth's gravitational field. STIRAP (Stimulated Raman
Processes) were used as beam-splitters, and the relative phase of the
atom traveling along the two arms of the interferometer was
measured. Because the interferometer was fixed to the Earth, this
phase difference is proportional to $g$. We propose a similar type of
experiment, but now in LEO where the interferometer is in free fall
\cite{SBD}. The term linear in $g$ now disappears, and the phase
difference will be proportional to the Riemann curvature tensor inside
the area encircled by two geodesics. 

Consider an atom interferometer in LEO at a distance
$R^{i}(t)$ from the center of the Earth. We choose a local
coordinate system $X^{\mu}$ fixed on the center-of-mass (CM) of the
apparatus. The signature of $g_{\mu\nu}$ is
$(-1,1,1,1)$; in linearized gravity $g_{\mu\nu}
=\eta_{\mu\nu}+h_{\mu\nu}$. Note also that
$|X^{i}|\ll|R^{i}|$; we expand $h_{\mu\nu}$ about $X^{i}=0$.

We then release an atom traveling along the orbit of the CM with
geodesic $\gamma_{1}$ (see Fig.~1). At $t_{A}$, STIRAP is used to
coherently split the atomic beam. One possible geodesic for the atom
is still $\gamma_{1}$; the other corresponds to a
\textit{different} geodesic $\gamma_{2}$. Note that the atom behaves
quantum mechanically, and that the superposition principle holds. 
It is \textit{not} possible to determine which geodesic any individual atom
will take. Because the spatial projection of both $\gamma_{1}$ and
$\gamma_{2}$ correspond to LEOs, they will intersect with one another
again after a time $T$ as shown in Fig.~2. Detectors can then be used
after coherent recombination at $t_B$ to determine the interference
pattern, and thus the phase shift $\Delta\phi$ that the atom picks up
between the two possible geodesics. The loss of fringe visibility would be
a measure of decoherence of the atom. Importantly, the combined
spacetime path $\gamma=\gamma_{1}\cup \gamma_{2}$ is closed (see
Fig.~1), and forms the boundary of a spacetime surface
$\mathcal{D}$. If the coherence time of the atom is too short to allow
the atomic beam to recombine due to drift, additional STIRAP beam
splitters can be used to force recombination. However, let us  
assume that the coherence time is long enough to allow Fig.~2.

The Schr\"odinger equation for the atom is
\begin{equation}
i\hbar\frac{\partial\psi}{\partial t}=-\frac{\hbar^{2}}{2m}\nabla^{2}%
\psi+i\hbar N_{i}\nabla^{i}\psi-mN_{0}\psi,\label{1}%
\end{equation}
where $\psi$ is the wavefunction, $m$ is the mass of the atom, and
using standard methods, 
\begin{align}
N_{0}  & =\frac{1}{4}X^{i}X^{j}\partial_{i}\partial_{j}h_{00}(R(t))+\frac{1}%
{2}X^{j}X^{k}\frac{dR^{i}}{dt}\partial_{j}\partial_{k}h_{0i}(R(t)),\nonumber\\
N_{i}  &
=X^{j}\partial_{j}h_{0i}(R(t))-\frac{1}{2}X^{j}X^{k}\partial_{j}\partial 
_{k}h_{0i}(R(t)).
\label{2}
\end{align}
Our results here hold only for stationary
metrics. However, from \cite{GLF} and \cite{ADS1995} our
result for $\Delta\phi$ given below holds as long as the local
curvature varies slowly in $X^{\mu}$. From \cite{GLF},
$N_\mu=(N_0,N_i)$ is the four-velocity field acting on the test
particle induced by the tidal field as seen by an observer at the CM.

An atom forms a wave-packet that propagates along either
$\gamma_{1}$ or $\gamma_{2}$. Taking the eikonal approximation
$\psi=e^{imS/\hbar}\psi_{0}$, where $\psi_{0}$ is the solution of
eq.~$(\ref{1})$ in the absence of $(N_{0},N_{i})$, 
\begin{equation}
\frac{\partial S}{\partial t}=-\frac{1}{2}|\nabla S|^{2}+N_{i}\nabla
^{i}S+N_{0},\qquad0=\frac{\psi_{0}}{2}\nabla^{2}S+(\nabla_{i}S-N_{i})\nabla^{i}\psi_{0}.
\label{3}
\end{equation}
From eq.~$(\ref{2})$, $\nabla_{i}N^{i}=0$. Thus, 
eq.~$(\ref{3})$ becomes
\begin{equation}
\frac{\partial S}{\partial t}=N_{0},\qquad\nabla_{i}S=N_{i},
\label{4}
\end{equation}
neglecting terms of $\mathcal{O}(N^{2})$. Solving eq.~$(\ref{4})$
\begin{equation}
S(X)=\int_{0}^{X^{\mu}}N_{\mu}d\tilde{X}^{\mu},\label{5}%
\end{equation}
integrated along $\gamma_{1}$ or $\gamma_{2}$. The exponentiated 
form of eq.~$(\ref{5})$ is Yang's nonintegrable phase factor \cite{Yang}, 
but with $N_{\mu}$ instead of the vector potential $A_{\mu}$.
$N_{\mu}$ \textit{does} pay a similar role in general relativity as
the gauge field arising from local Galilean invariance in the
nonrelativistic limit \cite{GLF}. Thus,
\begin{equation}
\Delta\phi=\frac{m}{\hbar}\int_{\gamma_{2}}N_{\mu}d\tilde{X}^{\mu}
-\frac{m}{\hbar}\int_{\gamma_{1}}N_{\mu}d\tilde{X}^{\mu}
=\frac{m}{\hbar} \int_{\mathcal{D}} R_{0i0j}(R(t))\tilde{X}^{i}
\> d\tilde{t}\> d\tilde{X}^j
\label{6}
\end{equation}
by Stokes' theorem. Since the normal to 
$\mathcal{D}$ is a spacelike vector, 
\begin{equation}
\Delta \phi\approx \frac{m}{\hbar} \vert R_{0i0j}\vert \mathcal{A}T
\label{6b}
\end{equation}
for $R_{0i0j}$ varying slowly in time. $\mathcal{A}$ is the area  
contained in the two intersecting orbits in Fig. 2, and $T$ is the
transit time. Notice that while $\Delta\phi$ is proportional to $R_{0i0j}$
and is thus independent of the choice of frame in the nonrelativisitc limit,
there is an unexpected dependence of $\Delta\phi$ of a test particle 
on its mass. For a cesium 
atom in a LEO of 480 km the phase difference is $\Delta \phi_{Earth}
\approx (mGM_\oplus/R^3\hbar)\mathcal{A}T\approx 0.26\>\>\mathcal{A}T$ 
rad cm$^{-2}$ s$^{-1}$.

The phase difference is proportional to the area inside the arms of 
the interferometer. Ultimately, this area is limited by the coherence
time of the atoms. Using the parameters in \cite{PCC} $T= 0.16$ s  
and $\mathcal{A}=0.01$ cm${}^{2}$, we get $\Delta\phi_{Earth}\approx
4\textrm{x} 10^{-4}$ rad. A single-measurement fringe resolution
$\approx 0.074$ rad has been demonstrated in Earth-based experiments
\cite{PCC}, and after one-minute's integration time, this resolution
improves to $0.011$ rad. Measurement of $\Delta\phi_{Earth}$  
requires the fringe sensitivity be increased by a factor
of 27 beyond the above. However, $\Delta\phi_{Earth}\sim T^3$
($\mathcal{A}\sim (vT)^2$ where $v$ is the velocity of the cesium atom
after STIRAP), and increasing the coherence time $T$ from $0.16$ s to
$0.5$ s will bring $\Delta\phi_{Earth}$ within the sensitivity of 
current interferometers.

That our proposed experiment is within the realm of feasibility is due
to the advances in atom interferometry. These interferometers are so
sensitive because beams of atoms are used instead of beams of
photons. At 100 GeV/c${}^{2}$, the rest-mass of an atom is larger than
the effective gravitational mass of a typical photon of 1 eV/c${}^2$
by eleven orders of magnitude. New interferometers based on
Bose-Einstein Condensates (BECs) in development \cite{TSKSK,
  SDECDHHRP} can increase the sensitivity of atom-based 
interferometry significantly. In-situ phase measurements of BECs
\cite{CHMW, HMWC, HDKTBEDJHRP} offer the potential of even more 
sensitive measurements of curvature. Moreover, with long coherence
times, vortices in BECs and superfluids may allow sensitive
interference measurements of curvature.  

Gravitational waves from astrophysical sources in the tens of
kilohertz range would cause a phase shift four orders of magnitude
smaller than $\Delta\phi_{Earth}$. This could in principle be detected
with future improvements in atom interferometry.

\begin{acknowledgments}
ADS and RYC were supported by a grant from the Office of Naval Research.
We also thank Stephen A. Fulling for critically reading the manuscript.
\end{acknowledgments}


\pagebreak

\begin{figure}[ptb]
\includegraphics[width=0.7\textwidth]{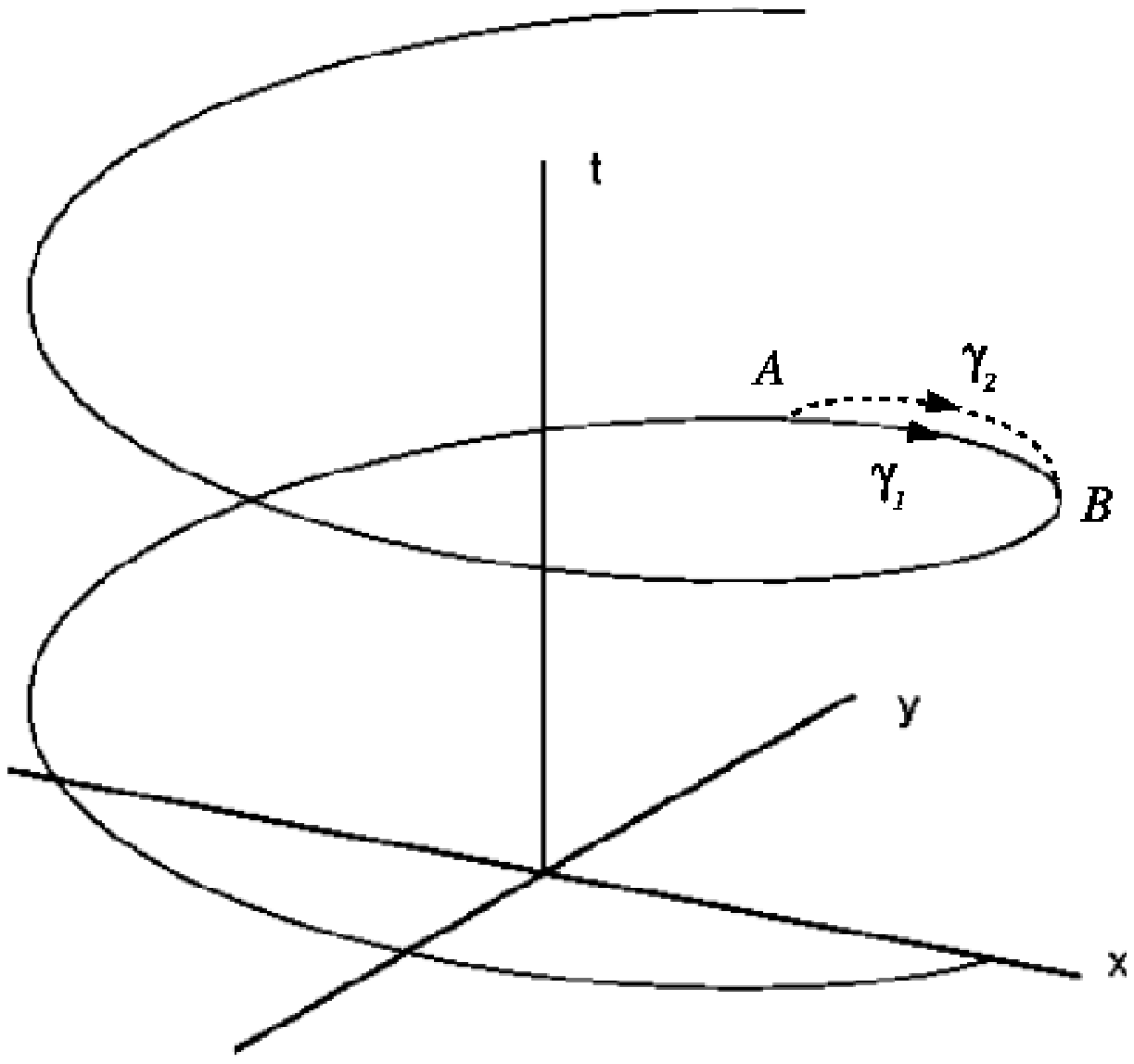} \caption{Sketch of the the
spacetime diagram of the interferometry measurement.}
\label{Fig-1}
\end{figure}

\pagebreak
\begin{figure}[ptb]
\includegraphics[width=0.9\textwidth]{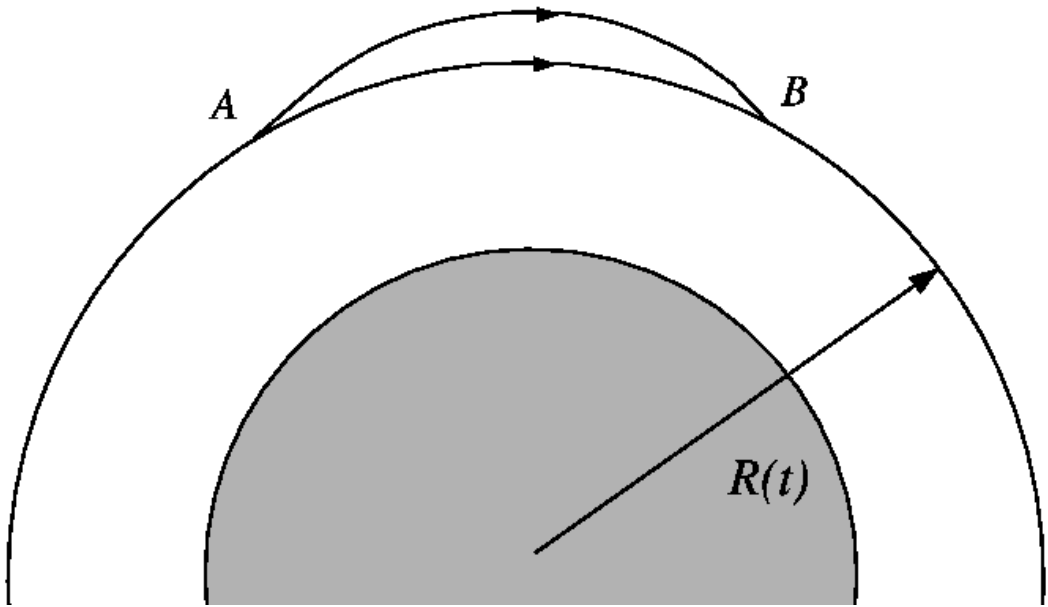} \caption{Sketch of the atomic
CM orbit around the Earth. An atomic beam is coherently split at $A$, and 
recombined coherently at $B$.}
\label{Fig-2}
\end{figure}
\end{document}